\documentclass[10pt,conference]{IEEEtran}

\usepackage[utf8]{inputenc}
\usepackage[T1]{fontenc}
\usepackage{amsmath,amssymb}
\usepackage{graphicx}
\graphicspath{{figures/}}
\usepackage{booktabs}
\usepackage{array}
\usepackage{capt-of}
\usepackage{xcolor}
\usepackage{url}
\usepackage[hidelinks]{hyperref}
\usepackage{listings}
\usepackage{xspace}
\usepackage{cite}
\usepackage{tikz}
\usetikzlibrary{arrows.meta,positioning,fit,backgrounds,shapes.multipart,calc,patterns,patterns.meta}

\newcommand{\coursename}{\mbox{The ABCs of Quantum Computing}\xspace}
\newcommand{\TIM}{TIM\xspace}

\begin{document}

% ============================================================
%  Title block
% ============================================================
\title{Using Learning Analytics to Study Novice Learners' Work on
Quantum Circuit Simulator Tasks}

\author{
\IEEEauthorblockN{Juuso Jaakola, Vlad Stirbu, Teiko Heinosaari}
\IEEEauthorblockA{\textit{Quantum Information and Computation Team}\\
Faculty of Information Technology\\
University of Jyväskylä, Finland\\
\{juuso.j.jaakola, vlad.a.stirbu, teiko.heinosaari\}@jyu.fi}
}

\maketitle

% ============================================================
%  Abstract
% ============================================================
\begin{abstract}
Interactive quantum circuit simulators are increasingly used to support introductory quantum computing education for learners from varied disciplinary backgrounds.
These environments can also generate detailed response logs of submitted circuits, but such data are rarely transformed into evidence that instructors can use for course redesign and development.

We present a learning analytics framework for quantum circuit simulator tasks and report its task-level branch on anonymised logs from an introductory quantum computing course delivered in the TIM learning environment.
The task branch characterises exercises through participation, completion, first-attempt accuracy, valid-attempt counts, sessionised active time, and observed answer-space reporting of common correct and incorrect constructions.
The dataset covers the 2024--2025 cohort of 492 learners, 47{,}533 valid submissions, and 21 obligatory quantum circuit exercises.
The framework identifies bottleneck tasks, broad solution spaces, and recurring wrong-answer patterns that are not visible from final scores alone.
Its most distinctive output is a plain-text representation of the observed answer space, designed to make open-ended circuit submissions directly interpretable to instructors.
We show how response logs can reveal actionable task-design issues and support concrete course redesign.
A learner-level branch is retained only as an ongoing research extension and is not a central empirical contribution of the present paper.
\end{abstract}

\begin{IEEEkeywords}

Quantum computing education; Learning analytics; Quantum circuit simulator;
Answer-space analysis; Data-informed course redesign
\end{IEEEkeywords}

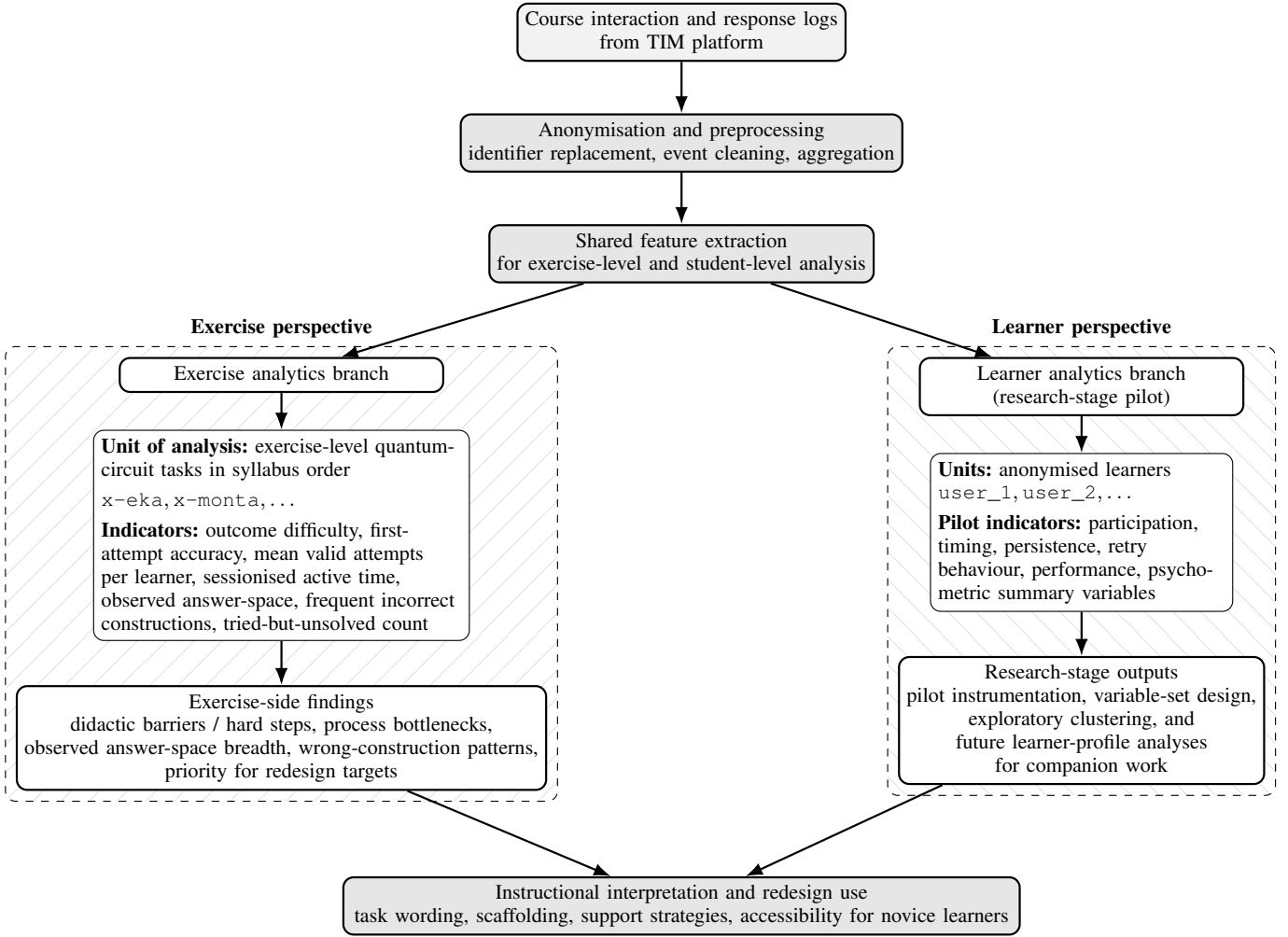
\begin{figure*}[t]
\centering
\resizebox{\textwidth}{!}{%
\begin{tikzpicture}[
    font=\footnotesize,
    >=Latex,
    node distance=6mm and 6mm,
    box/.style={draw, rounded corners, thick, align=center, inner sep=3pt, fill=white},
    smallbox/.style={draw, rounded corners, align=left, inner sep=3pt, fill=white},
    data/.style={box, fill=black!5, minimum width=44mm},
    process/.style={box, fill=black!10, minimum width=44mm},
    taskbox/.style={box, minimum width=44mm},
    learnerbox/.style={box, minimum width=44mm},
    tasksmall/.style={smallbox},
    learnersmall/.style={smallbox},
    note/.style={draw, dashed, rounded corners, inner sep=4pt, align=left},
    tasknote/.style={note, fill=black!1, pattern={Lines[angle=45,distance=8pt,line width=0.2pt]}, pattern color=black!12},
    learnernote/.style={note, fill=black!1, pattern={Lines[angle=135,distance=8pt,line width=0.2pt]}, pattern color=black!12}
]
\node[data] (source)
  {Course interaction and response logs\\from TIM platform};
\node[process, below=7mm of source] (anon)
  {Anonymisation and preprocessing\\identifier replacement, event cleaning, aggregation};
\node[process, below=7mm of anon] (split)
  {Shared feature extraction\\for exercise-level and student-level analysis};

\node[taskbox, below left=10mm and 6mm of split] (taskbranch)
  {Exercise analytics branch};

\node[tasksmall, below=5mm of taskbranch, text width=49mm] (taskmatrix)
  {\textbf{Unit of analysis:} exercise-level quantum-circuit tasks in syllabus order\\[1mm]
   $\texttt{x-eka}, \texttt{x-monta}, \ldots$\\[1mm]
   \textbf{Indicators:} outcome difficulty,
   first-attempt accuracy,
   mean valid attempts per learner,
   sessionised active time,
   observed answer-space,
   frequent incorrect constructions,
   tried-but-unsolved count};

\node[taskbox, below=6mm of taskmatrix] (taskfindings)
  {Exercise-side findings\\didactic barriers / hard steps, process bottlenecks,\\
   observed answer-space breadth, wrong-construction patterns,\\
   priority for redesign targets};

\node[learnerbox, below right=10mm and 6mm of split] (studentbranch)
  {Learner analytics branch\\(research-stage pilot)};
\node[learnersmall, below=5mm of studentbranch, text width=39mm] (studentmatrix)
  {\textbf{Units:} anonymised learners\\
   $\texttt{user\_1}, \texttt{user\_2}, \ldots$\\[1mm]
   \textbf{Pilot indicators:} participation, timing, persistence,
   retry behaviour, performance, psychometric summary variables};
\node[learnerbox, below=6mm of studentmatrix] (studentfindings)
  {Research-stage outputs\\pilot instrumentation, variable-set design,\\
   exploratory clustering, and\\
   future learner-profile analyses\\
   for companion work};
\node[process, below=20mm of $(taskfindings)!0.5!(studentfindings)$,
      minimum width=92mm] (synthesis)
  {Instructional interpretation and redesign use\\
   task wording, scaffolding, support strategies,
   accessibility for novice learners};
\draw[->, thick] (source) -- (anon);
\draw[->, thick] (anon) -- (split);
\draw[->, thick] (split) -- (taskbranch);
\draw[->, thick] (split) -- (studentbranch);
\draw[->, thick] (taskbranch) -- (taskmatrix);
\draw[->, thick] (taskmatrix) -- (taskfindings);
\draw[->, thick] (studentbranch) -- (studentmatrix);
\draw[->, thick] (studentmatrix) -- (studentfindings);
\draw[->, thick] (taskfindings) -- (synthesis);
\draw[->, thick] (studentfindings) -- (synthesis);
\begin{scope}[on background layer]
  \node[tasknote, fit=(taskbranch)(taskmatrix)(taskfindings),
        label=above:{\textbf{Exercise perspective}}] {};
  \node[learnernote, fit=(studentbranch)(studentmatrix)(studentfindings),
        label=above:{\textbf{Learner perspective}}] {};
\end{scope}
\end{tikzpicture}
}
\caption{Two-branch learning analytics and EDM pipeline for
TIM response logs. Both branches share the same source data and
preprocessing, but differ in unit of analysis and resulting
indicators. The task branch uses the 2024--2025 cohort
($N{=}492$) and provides the results reported in this paper. The
learner-branch pilot uses the 2025--2026 cohort ($N{=}209$) and is
included here only as a research-stage component of the broader
framework.}

\label{fig:pipeline_big}
\end{figure*}

% ============================================================
%  I. Introduction
% ============================================================
\section{Introduction}
\label{sec:intro}

Quantum computing is moving from graduate specialisation into

undergraduate teaching and open online courses faster than the educational research can keep up.
Interactive quantum circuit simulators have emerged as a standard pedagogical instrument in this transition~\cite{reinikainen2024qcforall,meyer2022interdisciplinary,goorney2024qctf}, because they make the otherwise abstract behaviour of gates and qubits more concrete.
The submission logs these simulators generate provide insight into students' submitted circuit constructions.
They do not record the intermediate steps of the construction process.
Even so, such logs still rarely translate into evidence that a course instructor can act on.
A recent systematic review of MOOC learning analytics~\cite{zhu2022mooc} identifies this as a field-wide gap.
Descriptive analytics studies are numerous, but studies that change teaching practice are not.

This paper addresses the feedback-loop side of that gap.
We describe a two-branch learning analytics pipeline built to support the data-driven redesign of \coursename, a large introductory quantum computing course for learners from mixed disciplinary backgrounds.
In the present paper, we report the task-level branch.
The learner-level branch remains an ongoing research component of the broader framework.
The pipeline is not presented as a finished product.
Instead, it is presented as a reproducible, shareable framework designed to evolve across successive course runs over a three-year horizon.

Quantum literacy is becoming relevant well beyond physics.
As quantum technologies emerge, a growing share of learners without a strong physics background will need a working understanding of quantum phenomena in computing.
Interactive circuit simulators help here.
They make the phenomenology more tractable without requiring the full Hilbert-space formalism at the outset.
Our pedagogical aim is therefore twofold: to improve this particular course and to develop analytic tools that may help other quantum computing courses reach and retain interdisciplinary learners.

To our knowledge no prior study has characterised the \emph{observed answer space} of open-ended quantum circuit construction tasks at scale using student submission logs.
We contribute:
\begin{enumerate}
    \item A two-branch analysis framework for an introductory quantum
    computing course. In the present paper, only the task-level branch
    is reported; the learner-level branch remains a research-stage
    extension for companion work.
    \item Systematic reporting of the observed answer space of each
    exercise, including observed correct constructions and the most
    frequent incorrect constructions, with representative answers
    rendered as ASCII schematics directly readable by instructors.
    \item A compact set of per-exercise indicators for
    \emph{outcome difficulty} and \emph{process difficulty}.
    Together, these indicators capture non-solution, first-attempt
    accuracy, retrying, and sessionised active time.
    \item Evidence from the 2024--2025 cohort that response-log
    analysis alone can surface actionable bottlenecks and
    instructional-design issues, and can already inform concrete
    course redesign actions.
\end{enumerate}

The rest of the paper reviews related work, describes the course and dataset, introduces the framework and key analytical definitions, reports the main findings, discusses limitations, and concludes with the redesign actions already taken and the broader implications of the framework.

% ============================================================
%  II. Background
% ============================================================
\section{Background}
\label{sec:background}

\subsection{Quantum computing education}
Introductory quantum computing courses have proliferated in the last five years~\cite{fox2020highered,meyer2022interdisciplinary,goorney2024qctf}.
These courses face a triple challenge: a mathematically demanding subject, a heterogeneous audience spanning physics, mathematics, computer science, and engineering, and the absence of the mature didactic traditions that exist for classical programming or introductory physics.
Misconception studies in introductory quantum mechanics~\cite{bouchee2022difficulties} document conceptual obstacles around superposition, measurement, and entanglement, but the body of work specifically on \emph{interactive circuit construction} tasks is thin.
\coursename course described here is delivered through the \TIM
learning environment~\cite{reinikainen2024qcforall} and uses an interactive drag-and-drop circuit simulator whose submission logs provide insight into students' constructed circuits, although the logs record only submitted solutions and not the intermediate steps of the construction process.

\subsection{Answer data in physics education research}
Physics education research (PER) has a long tradition of using student response distributions on concept inventories to identify consistent misconceptions, e.g.\ the Quantum Mechanics Concept Assessment~\cite{sadaghiani2015qmca}.
The methodological parallel with our work is direct: whereas PER analyses the distribution of selected distractors, we analyse constructed quantum circuits, which offer a different and somewhat more fine-grained view of students' submitted work and inferred reasoning.
To our knowledge such analyses have not been applied at scale to quantum circuit construction tasks.
Empirical insights into research-based teaching strategies in quantum education~\cite{donhauser2024empirical} underline both the appetite for such evidence and the current scarcity of it.

\subsection{Learning analytics and the redesign gap}
Romero and Ventura~\cite{romero2010edm} established the descriptive analytics framing that most educational data mining and learning analytics work has since followed~\cite{baker2014edm}.
Session-based engagement indicators are standard, and threshold-based sessionisation commonly uses inactivity cutoffs on the order of tens of minutes~\cite{kovanovic2015time}.
Log-derived indicators have also been linked to self-reported study strategies and to academic performance, which motivates using them as formative rather than purely descriptive evidence~\cite{gasevic2017strategies}.
The systematic MOOC-analytics review by Zhu et al.~\cite{zhu2022mooc} identifies the persistent gap we address here: descriptive findings rarely translate into redesign actions.

Although the course studied here is not a MOOC, its scale and its reliance on automatically logged submissions place it in the same methodological space, and we therefore draw our sessionisation practice from that literature.
Evidence from other online-learning contexts also suggests that even modest reflective prompts can improve learner behaviour~\cite{mitrovic2019nudges}. This supports also the broader use of instructional prompts to guide learning.
In the present course, response-log analysis informs revisions to hints and task wording to address common errors.

% ============================================================
%  III. Course and data
% ============================================================
\section{Course and data}
\label{sec:course-data}

\subsection{ \coursename course}
\label{sec:course}
The course is delivered through \TIM\ (The Interactive Material), an open-source, MIT-licensed learning platform developed and maintained by the Faculty of Information Technology at the University of Jyväskylä.\footnote{\url{https://tim.education/view/about/en-US}} TIM has been used since 2014 by more than 20{,}000 users across university and upper-secondary courses.
It supports document-based interactive material with built-in exercise components, automatic assessment, and course-level activity logging; the drag-and-drop quantum circuit simulator used here is implemented as one such component~\cite{reinikainen2024qcforall}.
\coursename is a three-module introductory quantum computing course (modules~A, B, and C) hosted on this platform.\footnote{\url{https://www.jyu.fi/en/it/introduction/kvanttilaskennan-aakkoset}} The course combines interactive reading material, a drag-and-drop circuit simulator, several obligatory circuit construction exercises, and conceptual single-answer questions.
Module~A introduces probability, binary arithmetic, and selected background ideas from quantum mechanics.
It also introduces the circuit simulator through classical gates such as X, CX, CCX, and SWAP, before moving on to the H, SX, and Z gates and the notion of inverse gates.
Module~B introduces matrix representations of gates and the mathematics of quantum information up to tensor products, without yet using complex numbers.
Module~C then extends the formalism to include complex numbers and goes deeper into the theory.
Across the three modules, the mathematical challenge increases toward the formal language of advanced quantum-information courses.
The course is offered in Finnish and in English.
It is open to enrol and delivered entirely online, but it is not a MOOC in the usual sense: roughly 60\% of enrolees take it for credit as part of a degree programme, and the circuit exercises are obligatory for them.
It therefore sits between an open online course and a conventional for-credit university course, and this mixed audience should be kept in mind when its completion figures are read.

\subsection{Dataset (2024--2025 cohort)}
\begin{table}[!t]
\caption{Dataset summary for the 2024--2025 task-branch cohort.}
\label{tab:dataset}
\centering
\begin{tabular}{lr}
\toprule
\textbf{Quantity} & \textbf{Value} \\
\midrule
Learners who submitted $\geq$1 valid attempt & 492 \\
Course completers (solved all obligatory exercises) & 397 \\
Non-completers & 95 \\
\quad of which mid-exercise non-completers & 48 \\
\quad of which clean-stop non-completers & 47 \\
Obligatory circuit construction exercises & 21 \\
Total valid submissions & 47{,}533 \\
\bottomrule
\end{tabular}
\end{table}

Table~\ref{tab:dataset} summarises the dataset.
Each circuit exercise awards 5 points for a fully correct submission and 0 otherwise; conceptual single-answer questions award 1 point.
Analyses in this paper restrict to the 21 obligatory circuit exercises.
We distinguish \emph{mid-exercise} non-completers (who left while still struggling with a partially solved exercise, $n{=}48$) from
\emph{clean-stop} non-completers (who stopped cleanly between
exercises, $n{=}47$).

Because learner work persists on the \TIM platform between academic years, some non-completers may later resume the course in a subsequent run.

\subsection{Ethics and data provenance}
Learners consent to participation upon course enrolment and can withdraw at any time.
The required data-protection documentation for anonymised research use has been submitted for review within the University of Jyväskylä.
Because the analyses in this paper are conducted on anonymised data, there is at this stage no separate ethics-committee approval reference attached to the study.
The manuscript therefore reports the current institutional review status directly, rather than implying a completed public approval record beyond what is presently available.
Under the University of Jyväskylä's research-data guidelines, anonymised log analysis of this kind does not require separate ethics-committee review, so the present status satisfies the requirements for publication.
The anonymised dataset may be made available after further review.

% ============================================================
%  IV. Framework
% ============================================================
\section{The two-branch framework}
\label{sec:framework}
The framework consists of two parallel branches operating on the same underlying submission log (Fig.~\ref{fig:pipeline_big}).
Each branch produces a primary indicator table and a set of visualisations.
The task-level branch also produces ASCII-rendered schematics of the observed answer space.

\subsection{Task-level branch}
For each of the 21 obligatory circuit exercises, the task-level branch computes a row of indicators spanning:
\begin{itemize}
    \item \emph{Participation and completion}: number of learners
    reaching, attempting, and solving.
    \item \emph{Outcome difficulty}: the fraction of learners who
    made at least one valid attempt on the exercise but did not
    eventually solve that exercise for full credit.
    Alongside this broader outcome signal we report \emph{drop-out
    hazard}, defined as the fraction of non-completers whose last
    valid course activity was that exercise.
    We use the latter term \emph{descriptively}: it is not intended
    as a formal survival-analysis hazard model.
    Drop-out hazard is the sharper redesign target, because it
    identifies where course exit actually concentrates.
    \item \emph{Process-difficulty signals}: first-attempt accuracy,
    mean valid attempts, and sessionised active time per learner.
    These describe different facets of cognitive load that can be
    isolated from log data alone; they are not proxies for each
    other.
    \item \emph{Answer space}: answer-space breadth (count
    of distinct correct constructions observed in submissions),
    and the top-$k$ most frequent observed correct and incorrect
    constructions themselves.
\end{itemize}
Outcome and process-difficulty indicators are collated in Section~\ref{sec:findings} as a single syllabus-ordered heatmap, with each indicator normalised within the course so that darker always means harder; reading across a row also exposes the distinct shapes of difficulty the course produces.

\subsection{Learner-level branch}
The learner-level branch is currently in a research and pilot phase within the broader two-branch framework.
Its purpose in the present paper is to document the intended future extension of the pipeline rather than to serve as a central empirical contribution.

When fully operationalised, the branch will produce a per-learner row of indicators grouped into four families: performance (scores, attempt counts, first-attempt success, accuracy), temporal effort (session counts, active working time, solving durations, inter-task latencies), path-related behaviour (revisits, average step distances, returns to earlier tasks), and questionnaire-based psychometrics (intrinsic motivation, flow, deliberate practice, achievement emotions; Table~\ref{tab:surveys}).

These indicators are being assembled for future learner-profile analysis.
The aim is to make the course audience more visible and to study variation in how learners engage with and progress through the course.
The first phase tests clustering under different variable sets drawn from these four indicator families.
It does not assume one fixed feature combination from the outset.
The goal is not to define stable learner types, but to identify patterns that may later inform instructional design and support strategies.
Table~\ref{tab:surveys} summarises the questionnaire instrument battery, which is specific to the learner-branch pilot for collecting psychometric data and is included here because those psychometric measures form one of the variable families considered in the pilot learner-branch design.

\begin{table}[!t]
\caption{Questionnaire battery administered in the learner-branch
pilot cohort ($N{=}209$, 2025--2026).}
\label{tab:surveys}
\centering
\footnotesize
\begin{tabular}{p{0.18\columnwidth}p{0.35\columnwidth}p{0.30\columnwidth}}
\toprule
\textbf{Time point} & \textbf{Instrument} & \textbf{Construct} \\
\midrule
Course entry (T0)
  & Grit-S~\cite{duckworth2009grits}
  & Grit (persistence / consistency) \\
  & SRL-O efficacy short scale (4 items)~\cite{broadbent2023srlo}
  & Online self-regulated learning efficacy \\
  & Adapted self-regulated learning strategy set (8 items)~\cite{barnard2009oslq}
  & Goal setting, monitoring, task strategies, time management, environment structuring, persistence \\
\midrule
Mid-course ($\sim$60\% complete)
  & IMI~\cite{mcauley1989imi}
  & Intrinsic motivation \\
  & Short Flow Scale
  & Flow experience \\
  & AEQ-S~\cite{bieleke2021aeqs}
  & Achievement emotions (enjoyment, anxiety, boredom) \\
  & Deliberate-practice pilot~\cite{ericsson1993deliberate}
  & Deliberate-practice tendencies \\
%\midrule
%End of course
  & SUS~\cite{brooke1996sus}
  & System usability of the synthesis of the \TIM environment and quantum circuit simulator \\
\bottomrule
\end{tabular}
\end{table}
In this paper, the learner-level branch remains a research-stage part of the broader framework.
Its first-phase exploratory clustering analyses, based on the 2025--2026 pilot cohort, are deferred to companion work.

\subsection{The answer-space ASCII visualiser}
\label{sec:answer-space-viz}
The most distinctive output of the task-level branch is the answer-space ASCII visualiser.
For each exercise, submissions are transformed into a common representation, allowing exactly identical submitted structures to be identified and counted together.

At this stage, we deliberately do not implement full canonicalisation or equivalence-class detection for logically equivalent circuits.
Because the analysis covers only 21 exercises, closely related constructions can still be inspected manually and, where appropriate, summarised by hand.
Preserving distinct submitted forms also keeps step-level variation visible.
In some exercises, that variation is itself pedagogically informative.

\begin{figure}[!t]
\centering
\includegraphics[width=0.95\columnwidth]{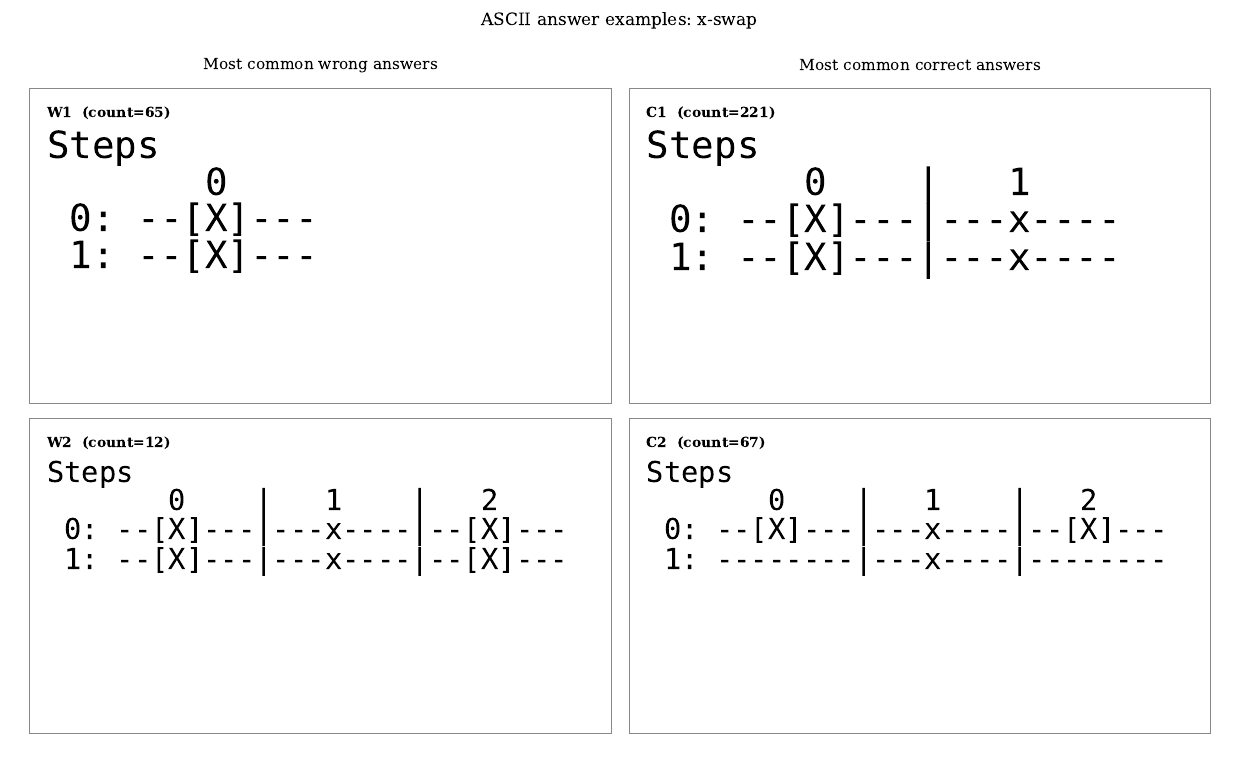}
\caption{Compact example of the answer-space ASCII visualiser for
\texttt{tehtava-x-swap}. The left column shows the two most common
incorrect submitted constructions and the right column the two most
common correct constructions, rendered as directly readable ASCII
circuit schematics.}
\label{fig:answer-space-xswap-ascii}
\end{figure}

In practice, the most frequent correct and incorrect constructions for each exercise are rendered as ASCII-style circuit diagrams that instructors can inspect directly without additional tools.
A compact example is shown in Fig.~\ref{fig:answer-space-xswap-ascii}.

The visualisation quickly surfaces two kinds of information.
A dominant incorrect construction may indicate a systematic misunderstanding, unclear task wording, or a simple recurring mistake.
Such patterns are useful both for redesign and for later PER-oriented misconception research.
Exercises with many distinct correct constructions reveal the diversity of valid strategies, including approaches that may be unexpected from the instructor's perspective.
That, in turn, helps when planning scaffolding, hints, and other support measures without privileging only one approach.

% ============================================================
%  V. Methods
% ============================================================
\section{Methods}
\label{sec:methods}

\subsection{What we can and cannot infer from logs}
Submission logs give us timestamps, scores, and the contents of each submitted circuit (and only the submitted circuit).
They do not give us pre-submission thinking time, off-screen consultation, attention, or self-reported understanding.
This limits what can be inferred from simple, commonly recorded answer-log data.

Our indicators therefore describe the cohort, not the individual learner.
They identify where work becomes hard and where learners leave, but cannot reconstruct the fine-grained interaction of any single person with the material.
Some courses do provide more detailed trace data, but such data are often highly specific to a particular platform or task design.
By contrast, the strength of the present approach lies in its reliance on simple answer logs, which are more likely to be available and comparable across future courses.

For a teacher with limited redesign time, this general analytical layer matters most.
It identifies which exercises are most likely bottlenecks (too hard, poorly worded, or insufficiently scaffolded) so that attention can be directed to the most immediate pain points.
Answer-space visualisation adds a complementary view: it shows aggregate behaviour as common stumbling blocks or recurring incorrect reasoning patterns that may indicate genuine misconceptions or, equally, language issues in the exercise material.

\subsection{Sessionisation and sessionised active time}
Sessions are constructed per (learner, exercise) pair by joining consecutive valid submissions whose timestamps lie within 15~minutes of each other.
The session ends at the first fully correct submission or at the last 0-point attempt plus a 15-minute tail.

The 15-minute threshold is drawn from the mainstream MOOC analytics range (15--60~minutes) reported by Kovanovi\'c et al.~\cite{kovanovic2015time}.
Because this choice is consequential for every time-based indicator, we report a sensitivity analysis at 10, 15, 20, and 30~minutes in Section~\ref{sec:sensitivity}.

It is important to note that the resulting sessionised active time is a heuristic estimate based on inter-submission intervals.
It does not directly measure actual time-on-task.
Instead, it is an indirect measure that supports comparative analysis of time investment across exercises and learners.
The difficulty of recovering true time-on-task from coarse activity traces is well documented in computing-education work on keystroke-level programming logs~\cite{hart2025time}, which reinforces our treatment of this indicator as comparative rather than absolute.

\subsection{Observed answer space}

For each exercise we report the set of distinct submitted constructions that received full credit, together with the most frequent incorrect constructions.
Two submissions are grouped as the same observed construction when their normalised submission strings (parsed from the \texttt{answers\_raw} JSON and sorted by \texttt{(time, name, targets, controls)}) are identical.
We do not currently detect equivalence between logically equivalent circuits that differ in the ordering of commuting gates.
Preserving such differences keeps step-level variation visible and is acceptable at the 21-exercise scale, but it does inflate breadth on multi-step exercises (see Section~\ref{sec:limitations}).

This summary reports observed forms, not all mathematically possible ones.
It highlights broad solution diversity, unexpected correct strategies, and recurring incorrect constructions that may reflect unclear wording, insufficient scaffolding, or common misconceptions.

\subsection{Outcome and process-difficulty indicators}
We separate outcome difficulty from process-difficulty indicators.
Outcome difficulty for exercise $e$ is the fraction of learners who made at least one valid attempt on $e$ but did not eventually solve it.
Drop-out hazard is the narrower signal: the fraction of non-completers whose last valid course activity was $e$.
We use ``hazard'' descriptively, not as a survival model.
Drop-out hazard is the sharper redesign target because it locates where course exit concentrates, while outcome difficulty captures eventual non-solution more broadly.
In the present cohort, clean-stop exits spread thinly across many exercises, while mid-exercise exits cluster on a small number of bottleneck tasks, chiefly \texttt{yhteenlasku-A} and
\texttt{tehtava-virheenkorjaus}.
The process-difficulty indicators---first-attempt accuracy ($\mathrm{faa}_e$), mean valid attempts ($\mathrm{att}_e$), and sessionised active time ($\mathrm{t}_e$)---describe distinct facets of cognitive load.
For reporting, first-attempt accuracy is shown in transformed form as first-try difficulty $(1-\mathrm{faa}_e)$ so that higher values consistently indicate greater difficulty.
These indicators are reported together in Section~\ref{sec:findings} as a per-exercise heatmap.

% ============================================================
%  VI. Preliminary findings
% ============================================================
\section{Preliminary findings}
\label{sec:findings}

The findings below describe one cohort and one course run.
We therefore focus on a small set of figures chosen for explanatory value.
We begin with a course-level progression view and then move to exercise-level indicators and answer-space findings.

Three results matter most for the present paper.
First, the exercise-level indicators identify a small set of bottleneck tasks and show that process difficulty is not captured by completion alone.
Second, observed answer-space reporting reveals design problems that aggregate outcome measures can hide; the clearest case is
\texttt{tehtava-virheenkorjaus}, where the breadth of correct
constructions and the recurrence of the unchanged starting circuit on the incorrect side showed that many learners had reconstrued the task more broadly than intended.
Third, these response-log findings are not only descriptive: they have already informed concrete redesign actions in the following course run, including revised wording and targeted hints.

\subsection{Course progression and bottlenecks}
Of the 492 learners who submitted at least one valid attempt, 397 completed all 21 obligatory circuit exercises (completion rate 80.7\%).
Of the 95 non-completers, 48 stopped mid-exercise and 47 stopped cleanly between exercises.
The clean-stop split indicates that exercise-level frustration is not the only mechanism behind course exit.
As noted in Section~\ref{sec:course}, this figure should not be read against open MOOC completion rates, which are typically far lower and vary widely across courses~\cite{jordan2015completion,guo2014mooc}; the more informative comparison would be to other for-credit introductory courses, which we leave for cross-institutional work.

\begin{figure}[tbp]
\centering
\includegraphics[width=0.95\columnwidth]{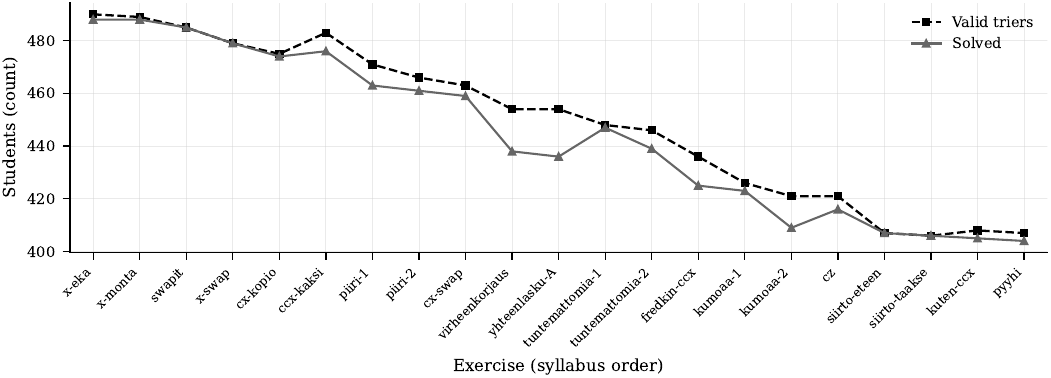}
\caption{Course progression by exercise in syllabus order. The solid
line shows the number of learners who made at least one valid attempt
on each exercise, and the dashed line shows the number who eventually
solved it. Local drops indicate where the cohort thins out, and the
separation between the curves shows outcome difficulty in count form:
learners who reached an exercise but did not solve it. The figure
therefore provides the first course-level view of bottlenecks before
the more local exercise-level plots.}
\label{fig:survival}
\end{figure}

Figure~\ref{fig:survival} provides the course-level overview of where the cohort thins out and where non-solution begins to accumulate exercise by exercise.

\subsection{Outcome and process indicators}
\label{sec:four-plots}
\begin{figure*}[tbp]
\centering
\includegraphics[width=1.00\textwidth, trim={0.4cm 1.1cm 0.8cm 1.0cm}, clip]{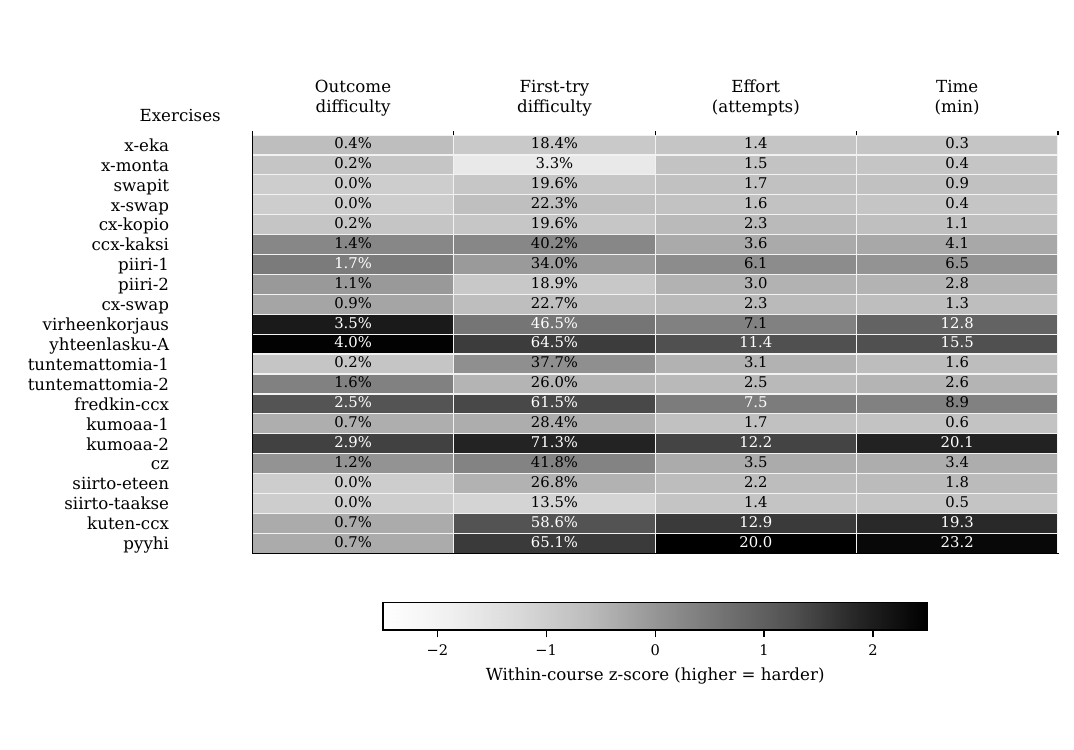}
\caption{Per-exercise difficulty indicators for the 21 obligatory
circuit exercises in syllabus order (top to bottom). The columns are
outcome difficulty (fraction of valid triers who never solved),
first-try difficulty (1 - first-attempt accuracy), mean valid attempts
per learner, and mean sessionised active time per learner. Cell
shading is the within-course $z$-score of each column, oriented so
that darker always indicates greater difficulty; cell annotations give
raw values in the units named in the headers. The figure is a
comparative summary across these 21 exercises, not an absolute
difficulty scale. Outcome difficulty spans a narrow absolute range
(0.0--4.0\%), so shading in that column reflects small differences;
the substantive contrast in this cohort is in the process indicators.}
\label{fig:indicator-heatmap}
\end{figure*}

Figure~\ref{fig:indicator-heatmap} collates all four indicators across the 21 exercises.
It is the exercise-level counterpart of the course-level gap in Fig.~\ref{fig:survival}, but resolved by indicator: reading down a column locates the hardest exercises on that dimension, and reading across a row shows whether an exercise is broadly demanding or hard in one specific way.

Two patterns stand out.
First, outcome difficulty is uniformly low --- even the hardest exercise leaves only 4\% of valid triers unsolved --- so eventual non-solution is not where the course's difficulty concentrates.
The substantive contrast is in the process indicators.
Second, mean valid attempts is the clearest process-difficulty signal in this cohort: a small set of exercises (\texttt{pyyhi},
\texttt{kuten-ccx}, \texttt{kumoaa-2}, \texttt{yhteenlasku-A})
requires an order of magnitude more iterative effort than its neighbours, even though nearly all learners eventually succeed.
Sessionised active time tracks the attempt count closely and adds temporal information rather than an independent signal.
First-try difficulty rises through the course but less sharply than effort, indicating that the later exercises are not so much harder to begin as harder to bring to a correct finish.

Reading across a row turns the same table into a difficulty typology.
A row that is dark on every column marks a broadly demanding exercise such as \texttt{yhteenlasku-A}; a row dark only in the effort and time columns marks an effort-heavy but conceptually shallow task such as
\texttt{pyyhi}; and a row dark only in first-try difficulty marks an
exercise that mainly resists the first attempt while resolving quickly afterwards.
For an instructor, this distinction is a fast guide to the kind of intervention an exercise needs: broad scaffolding, reduced iteration, or clearer wording.

\subsection{Long-tail effort distributions}
\begin{figure*}[tbp]
\centering
\includegraphics[width=\textwidth]{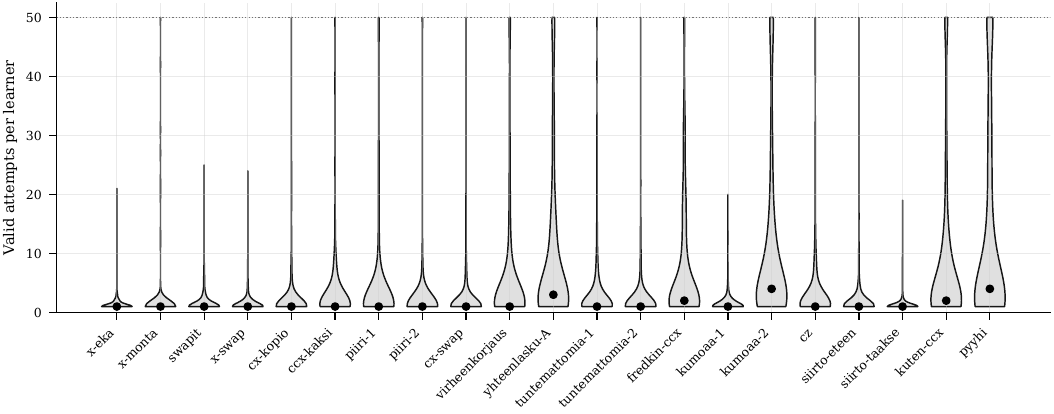}
\caption{Distribution of valid attempts per learner per exercise in
syllabus order, shown as violin plots with the display capped at 50
attempts. Most exercises have a dense low-attempt region together with
a long thin upper tail. The black dots mark the median number of valid
attempts for each exercise. This indicates that the typical learner solves
with moderate effort, while a smaller subset experiences persistent
difficulty. The figure is therefore most useful for revealing where
targeted scaffolding may be needed rather than for characterising the
average learner alone.}
\label{fig:submission-violins}
\end{figure*}

The mean-based indicators above hide within-exercise variance.
Figure~\ref{fig:submission-violins} shows that across most exercises the attempt-count distribution is dense at low values with a long thin upper tail: the typical learner solves with moderate effort while a smaller subset experiences persistent difficulty.
This distinguishes exercises that need targeted scaffolding for a struggling minority from those that are uniformly hard.

\subsection{Observed answer space: the \texttt{virheenkorjaus} redesign case}
\label{sec:virheenkorjaus}
The clearest redesign case in the present material is
\texttt{tehtava-virheenkorjaus}. The task was intended as a small
correction to a pre-filled circuit, but the observed answer space shows that many learners approached it much more broadly.
This is precisely the kind of issue that pass rates alone do not make visible.

\begin{figure}[!b]
\centering
\includegraphics[width=\linewidth]{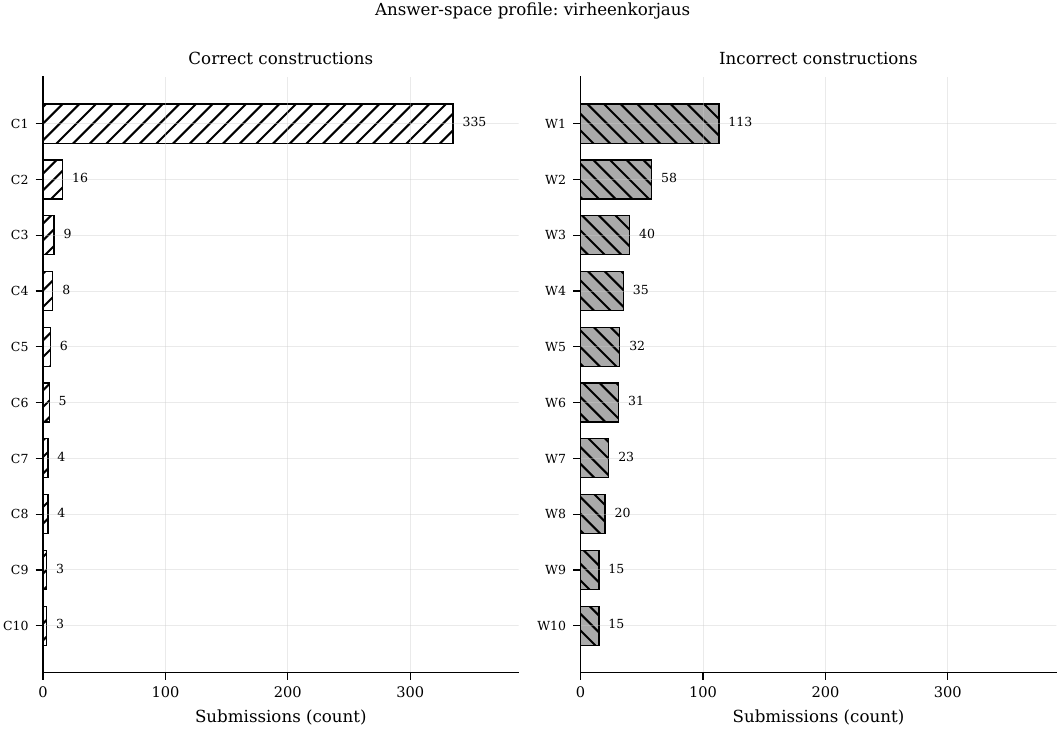}
\caption{Observed answer-space profile for the first ten correct and
incorrect submitted constructions in \texttt{tehtava-virheenkorjaus}.
The two panels summarise the most frequent observed correct and
incorrect forms in the submitted data. The figure shows that the intended solution accounted for the clear majority of correct submissions, but the tail of less frequent correct constructions started with low count and extended unexpectedly far.}
\label{fig:answer-space-profile}
\end{figure}
The answer-space profile suggests that the exercise did not function primarily as a narrow error-correction task.
Instead, many learners appear to have treated it as a more open-ended construction problem.
This points to an instructional-design issue rather than simply a hard conceptual step.

\begin{figure}[!b]
\centering
\scriptsize
\setlength{\tabcolsep}{2.5pt}
\renewcommand{\arraystretch}{1.05}
\begin{tabular}{@{}rr>{\ttfamily\raggedright\arraybackslash}p{0.60\columnwidth}@{}}
\toprule
Rank & Count & Correct answers (canonical JSON, truncated) \\
\midrule
1  & 335 & \detokenize{... "controls":[3,4],"targets":[7],"time":5 ...} \\
2  & 16  & \detokenize{... "controls":[4,5,6],"targets":[7],"time":5 ...} \\
3  & 9   & \detokenize{... "controls":[3,5],"targets":[7],"time":5 ...} \\
4  & 8   & \detokenize{... "controls":[4,5,6],"targets":[3],"time":4 ...} \\
5  & 6   & \detokenize{... "controls":[3,6],"targets":[7],"time":5 ...} \\
6  & 5   & \detokenize{... "controls":[0,1],"targets":[5],"time":2 ...} \\
%7  & 4   & \detokenize{... "controls":[0,1],"targets":[6],"time":2 ...} \\
%8  & 4   & \detokenize{... "controls":[0,2],"targets":[7],"time":2 ...} \\
%9  & 3   & \detokenize{... "controls":[3,4,5],"targets":[7],"time":5 ...} \\
%10 & 3   & \detokenize{... "controls":[3,4,6],"targets":[7],"time":5 ...} \\
%11 & 3   & \detokenize{... "controls":[3,5],"targets":[7],"time":5 ...} \\
%12 & 3   & \detokenize{... "controls":[1,2],"targets":[7],"time":2 ...} \\
%13 & 3 & \detokenize{... "controls":[1,2],"targets":[7],"time":2 ...} \\
%14 & 2 & \detokenize{... "controls":[4,5,6],"targets":[3],"time":4 ... } \\
%15 & 2 & \detokenize{... "controls":[],"targets":[2],"time":5 ... } \\
%16 & 2 & \detokenize{... "controls":[],"targets":[7],"time":5 ... } \\
%17 & 2 & \detokenize{... "controls":[0,1],"targets":[5],"time":2 ... } \\
\midrule
\multicolumn{3}{c}{\emph{\dots 45 further distinct correct forms omitted \dots}} \\
\midrule

52 & 1   & \detokenize{... "targets":[7],"time":0 ... "controls":[4],"targets":[7],"time":6 ...} \\
53 & 1   & \detokenize{... "controls":[1],"targets":[3],"time":0 ... "targets":[7],"time":6 ...} \\
54 & 1   & \detokenize{... "controls":[],"targets":[3],"time":0 ... "controls":[5,6],"targets":[7],"time":6 ...} \\
\bottomrule
\end{tabular}
\caption{Distinct correct submitted constructions for
\texttt{tehtava-virheenkorjaus} from teacher aimed report. Although the exercise was framed as a
small correction task, the submitted data contained 54 distinct
correct constructions. The table shows the highest-frequency forms and
the final three ranked variants; canonical JSON strings are truncated
for readability.}
\label{fig:answer-space-table}
\end{figure}

Taken together, Figs.~\ref{fig:answer-space-profile} and
\ref{fig:answer-space-table} show why systematic reporting of the
observed answer space is useful.
Most of these learners did eventually reach a correct answer, so the issue would remain largely invisible in completion statistics alone.
The answer-space view turns that hidden design problem into something concrete enough to revise.

\begin{figure*}[p]
\centering
\includegraphics[width=0.88\textwidth]{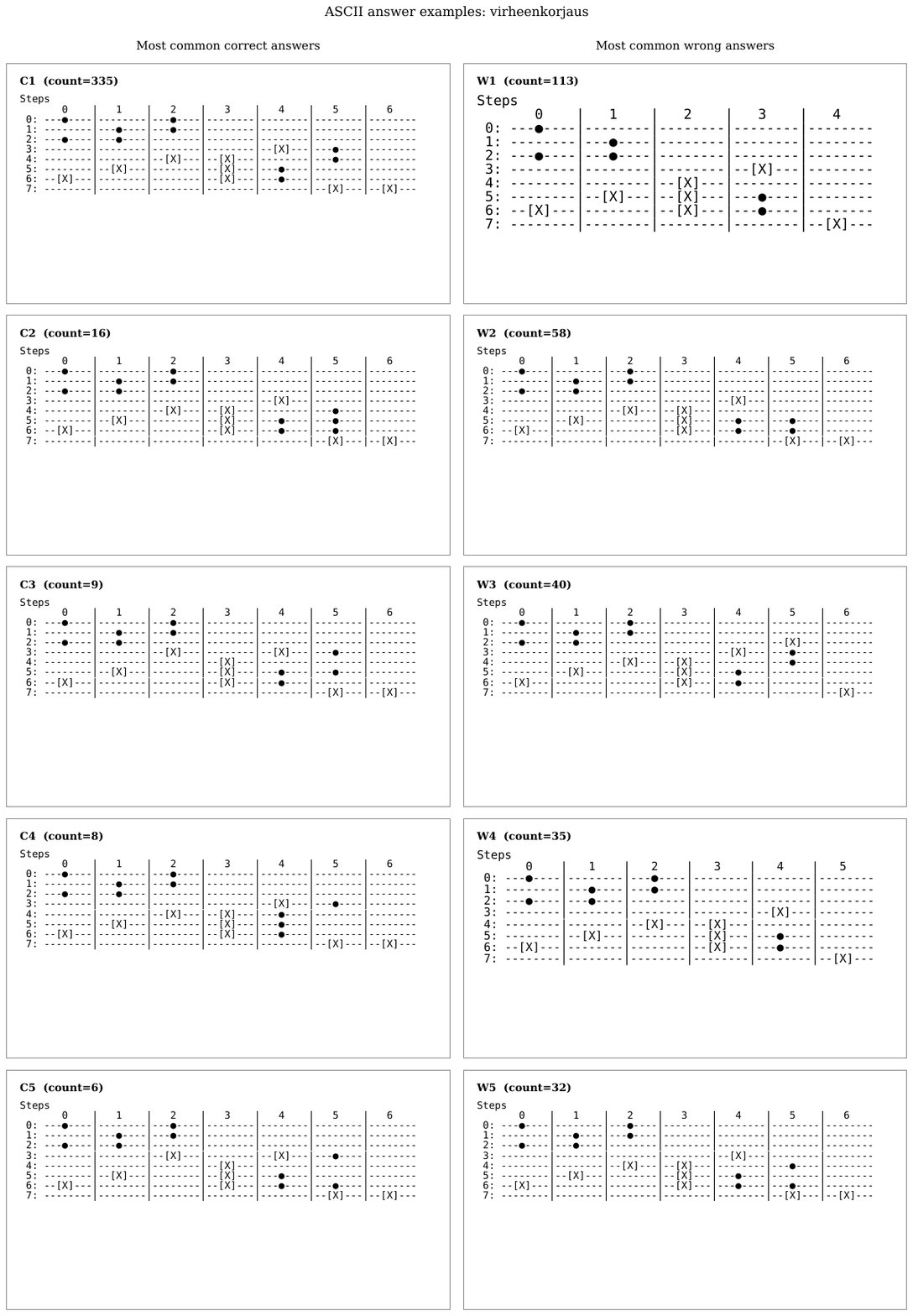}
\caption{Full-page ASCII rendering of representative observed answers
for \texttt{tehtava-virheenkorjaus}. This larger view retains the
broader set of dominant submitted forms for visual inspection while the
main text uses the ranked-table view in Fig.~\ref{fig:answer-space-table}.}
\label{fig:answer-space-ascii-full}
\end{figure*}

\subsection{Sensitivity to the session-gap threshold}
\label{sec:sensitivity}
Table~\ref{tab:sensitivity} reports a simple sensitivity check in which the session-gap cap is varied across 10, 15, 20, and 30~minutes while the closure tail is kept fixed at 15~minutes.
Each alternative cap is compared against the 15-minute reference.

\begin{table}[!t]
\scriptsize
\setlength{\tabcolsep}{2.5pt}
\caption{Session-gap sensitivity with a fixed 15-minute closure tail.
Values are compared with the 15-minute reference. Spearman
$\rho$ reports whether the exercise ordering changes, and
$\Delta$ reports how much the indicator values change on average.
For mean sessionised active time and mean sessions, $\Delta$ is the
mean absolute percentage change; for time difficulty, it is the mean
absolute change in $z$-score units.}
\label{tab:sensitivity}
\centering
\begin{tabular}{lcccccccc}
\toprule
\multicolumn{1}{c}{} & \multicolumn{2}{c}{10 m} & \multicolumn{2}{c}{15 m} & \multicolumn{2}{c}{20 m} & \multicolumn{2}{c}{30 m} \\
\cmidrule(lr){2-3}\cmidrule(lr){4-5}\cmidrule(lr){6-7}\cmidrule(lr){8-9}
Indicator & $\rho$ & $\Delta$ & $\rho$ & $\Delta$ & $\rho$ & $\Delta$ & $\rho$ & $\Delta$ \\
\midrule
Mean sessionised active time & 1.00 & 1.31\% & 1.00 & 0.00\% & 1.00 & 0.71\% & 1.00 & 3.62\% \\
Mean sessions & 0.99 & 2.76\% & 1.00 & 0.00\% & 1.00 & 1.67\% & 1.00 & 3.08\% \\
Time difficulty ($z$) & 1.00 & 0.00 & 1.00 & 0.00 & 1.00 & 0.00 & 1.00 & 0.01 \\
\bottomrule
\end{tabular}
\end{table}
The exercise ordering is essentially unchanged across the tested caps, and the corresponding value changes remain small when the closure tail is held fixed at 15~minutes.

% ============================================================
%  VIII. Limitations
% ============================================================
\section{Limitations}
\label{sec:limitations}

\paragraph{One cohort, one course}
All findings are drawn from the 2024--2025 run of one course at one institution.
Cross-cohort replication on the 2025--2026 run is in progress.
External validity to other \TIM-hosted science courses is plausible but unverified.

\paragraph{Log data describes the cohort, not the individual}
Submission logs support only coarse inferences about how learners interacted with the material.
We do not observe pre-submission thinking time or off-screen consultation; ``sessionised active time'' is an inter-submission proxy with a capped 15-minute gap (sensitivity check in Section~\ref{sec:sensitivity}).
Our findings describe the cohort at a general level: where work becomes hard, where learners leave, and what patterns recur.
They do not reconstruct any individual learner's interaction precisely.

\paragraph{Current reporting normalisation does not account for gate commutativity}
Two circuits that differ only in the order of commuting gates are currently counted as distinct observed constructions.
This inflates observed answer-space breadth on exercises with many independent time steps.
A commutativity-aware extension of the grouping is in progress.

\paragraph{Qubit-placement error analysis was attempted and withdrawn}
An earlier pipeline version included a per-exercise heatmap of (intended qubit, submitted qubit) errors, intended to surface systematic qubit-indexing confusion.
On testing we concluded that the approach cannot reliably distinguish genuine indexing errors from benign multi-solution diversity: for exercises with many valid constructions, the ``intended'' reference is an arbitrarily chosen most-common correct answer, which marks alternative-order solutions as wrong.
The analysis has been withdrawn from the framework.
It is retained only for internal diagnostic use on simple single-gate, single-step exercises.

As a concrete example of the type of design flaw that remains visible in the retained framework, Figs.~\ref{fig:answer-space-profile},
\ref{fig:answer-space-table}, and \ref{fig:answer-space-ascii-full}
show that \texttt{tehtava-virheenkorjaus}, intended as a simple ``find the error and correct it'' exercise, elicited an unusually broad observed answer space.
The most common wrong pattern was a trivial submission with no change to the circuit.
The correct side contained 54 distinct submitted constructions.
This suggests that some learners reconstrued the task as a broader circuit-building problem rather than a local correction, motivating clearer wording and updated hints to reduce avoidable frustration.

% ============================================================
%  IX. Redesign actions and conclusion
% ============================================================
%\clearpage
\section{Redesign actions and conclusions}
\label{sec:redesign}
\label{sec:conclusion}
A distinguishing feature of this work is that the analyses have already driven concrete changes to the course.
Between the 2024--2025 and 2025--2026 runs, an unrestricted hint panel was added below every circuit exercise; learners can open it at any point with no scoring penalty.
In addition, the two bottleneck exercises identified by the answer-space analysis received hints that targeted the candidate misconceptions surfaced by the wrong-construction profiles.
The wording of \texttt{tehtava-virheenkorjaus} was revised to clarify the intended scope of the correction task.

The two-branch pipeline reported here covers 492 learners, 47{,}533 valid submissions, and 21 obligatory circuit simulator exercises from the 2024--2025 cohort.
The central contribution of the present paper is to show that response-log analytics alone can support actionable task-level redesign.
Its most distinctive output is the systematic reporting of the observed answer space of open-ended quantum circuit construction tasks.
This output is made directly legible to instructors through ASCII-rendered representative constructions and compact outcome- and process-difficulty indicators.

The learner-level branch remains an ongoing research-stage component of the broader framework.
Its first-phase exploratory clustering analyses, based on the 2025--2026 pilot cohort and alternative variable sets, are deferred to a forthcoming companion paper.
Cross-cohort replication of the task branch and a commutativity-aware extension of the answer-space grouping are in progress for the 2025--2026 run.
The analysis pipeline is being prepared for open-source release alongside the companion paper.

The broader aim is to develop a shareable toolkit for \TIM-hosted science courses and, more broadly, to support quantum technology literacy beyond physics.

% ============================================================
%  Acknowledgements
% ============================================================
\section*{Acknowledgements}
This work was supported by the Finnish Ministry of Education and Culture through the Quantum Doctoral Education Pilot Program (QDOC VN /3137/2024-OKM-4) and by the Research Council of Finland through the Finnish Quantum Flagship project (359240).

\emph{AI tooling disclosure:}
Portions of the analysis pipeline code were authored, reviewed, refactored, and locally optimised with assistance from large language models, including Claude Sonnet 4.6 (Anthropic) and ChatGPT 5.5 (OpenAI).
Grammar checks for the manuscript language were similarly supported.
All analytical choices, data interpretation, and manuscript content are the authors' own, and every LLM-assisted change to code and text was reviewed by the authors before inclusion.

% ============================================================
%  Bibliography
% ============================================================
\bibliographystyle{IEEEtran}
\bibliography{references}

\end{document}